# Role of Interstitial Pinning in the Dynamical Phases of Vortices in Superconducting Films

R. P. Simões, P. A. Venegas and D. F. Mello

Faculdade de Ciências, Unesp-Universidade Estadual Paulista, Departamento de Física, Bauru, SP, CP 473, 17033-360, Brazil

**Abstract** We analyze the vortex dynamics in superconducting thin films with a periodic array of pinning centers. In particular, we study the effect of anisotropy for a Kagomé pinning network when longitudinal and transverse transport currents are applied. By solving the equations of motion for the vortex array numerically at zero temperature we find different phases for the vortex dynamics depending on the pinning and driving force. An unusual sequence of peaks along and perpendicular to the main hexagonal lattice axes is observed for the differential resistance reflecting the anisotropy of the transport properties and the complex behavior of the vortex system. This behavior may be understood in terms of interstitial pinning vacancies, which create channels of vortices with different pinning strength.

## 1. INTRODUCTION

The study of current-driven vortex lattices in type-II superconductors has attracted great interest from both experimental and theoretical point of view [1-13]. When the Vortex lattice is forced to move by applying a transport current in a model with pinning sites that can trap vortices, the movement becomes complex and an analytical solution of the problem is not possible. In particular, the Kagomé structure has been extensively used to study several unique features of this type of pinning lattice [3]. Besides, experiments using triangular pinning have been carried out recently that used different transport current directions and found a strong anisotropy for the transport properties [4]. We performed molecular dynamics simulations to analyze the different vortex dynamic phases for a thin type II superconducting film, infinite in $x$ and $y$ directions, with Kagomé pinning network. In particular, we calculate the vortex trajectory, differential resistance and structure factor for transport currents applied longitudinally ($y$) and transversally ($x$). The matching current for each dynamic phase transition is accurately determined by the differential resistance peaks. Our analysis for the Kagomé pinning lattice shows an unexpected sequence of peaks in the differential resistance, revealing that the interstitial pinning centers play a crucial role for the anisotropy of the dynamic phases and create channels of vortices with different matching transport currents.

## 2. MODEL

The equation of motion of a vortex at the position $\mathbf{r_i}$ at zero temperature on a triangular lattice is given by:

$$\frac{d\mathbf{r}_i}{dt} = -\sum_{j \neq i} \nabla_i U_{vv}(r_{ij}) - \sum_p \nabla_i U_p(r_{ip}) + \mathbf{F}_c \quad (1)$$

where $U_{vv}(r_{ij})$ is the vortex-vortex interaction:

$$U_{vv}(r_{ij}) = C_v \ln(r_{ij}) \quad (2)$$

$r_{ij}$ is the vortex-pinning distance and $C_v$ the strength of the vortex-vortex interaction, defined as $C_v = \Phi_0^2/8\pi\Lambda$ (with $\Lambda$ the effective penetration depth and $\Phi_0$ the flux quantum). The vortex-pinning interaction, $U_p(r_{ip})$, is:

$$U_p(r_{ip}) = -C_p e^{-(r_{ip})^2} \quad (3)$$

with $r_{ip}$ is the vortex-pinning distance, $C_p$ the strength of the vortex-pinning interaction. The last term in Eq. (1), $\mathbf{F}_c$, is the driving force associated with a transport current $\mathbf{J}$:

$$\mathbf{F}_c = \frac{\Phi_0}{c} \mathbf{J} \times \hat{\mathbf{z}} \quad (4)$$

Figure 1 shows a hexagonal vortex lattice with pinning centers located on a sublattice forming a Kagomé array. The ratio of number of vortices to the number of pinning centers is $n_v/n_p = 4/3$. Note, due to the super-imposed Kagome lattice for the pinning centers the density of free vortices alternates between adjacent lines in x and y direction forming so-called channels along which the vortices move. For longitudinal direction ($y$), the ratio of the number of vortices to pinning centers switches between 1/1 and 2/1. For transverse direction ($x$), we observe the same ratios but different symmetry. Despite even rows have the same pinning density, the pinning centers of alternate even rows are shifted. This intricate topology is the origin of the anisotropy and the complex dynamic vortex behavior.

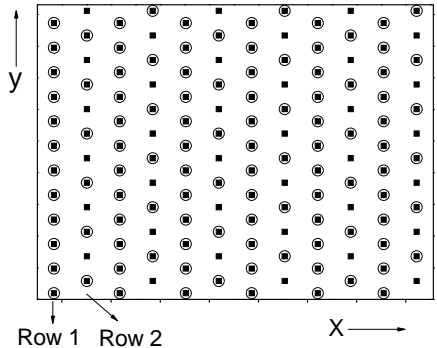

**Figure 1.** The black dots represent the vortices and the open circles the pinning centers.

In our molecular dynamics simulation we use a box of size $L_x$ by $L_y$, and periodic boundary conditions to simulate the infinite size of the sample. We consider a system with 144 total vortices. We start the simulation process by relaxing the vortex lattice without a transport current to find a stable equilibrium configuration, and then we use this as the initial boundary condition. Each simulation is started at $F_c$=0.0 and the force is increased in steps of $\Delta F_c$=0.05 increments up to values as high as $F_c$=4.0. For each value of force we take 2,000 normalized time steps for equilibration and 30,000 time steps for evaluation of the time averages [5]. The length scales are normalized by $4\xi$, the energy scales by $A_v$, and the time by $16\eta\xi^2/A_v$, with $\xi$ the coherence length and $\eta$ the Bardeen-Stephen friction.

To analyze the vortex dynamics we use a fixed vortex density $n_v$=0.12, an applied magnetic magnetic field $H=4/3B_\Phi$ along the z-axis, and a transport force in $x$ or $y$ direction. $B_\Phi$ is the first matching field defined as the field when the number of vortices equals the number of pinning sites. The different vortex regimes are characterized by calculating the vortex trajectories, the time average of the vortex velocity, structure factor, transverse diffusion and the vortex velocity derivative $dV/dF_c$, which is proportional to the differential resistance $dV/dF_c=\rho_f dE/dJ$ ($\rho_f$ is the flux flow resistivity) [5].

## 3. RESULTS AND DISCUSSION

In the following, we show molecular dynamics simulations of the vortex system for Kagomé pinning lattice and applying a driving force into longitudinal and transvers direction.

Figure 2 depicts the differential resistance for a force applied in longitudinal direction showing two peaks. From the analysis of the trajectories of the vortices and the structure factor, we can deduce that the two peaks correspond to phase transitions between the three different dynamical vortex regimes: I) For forces smaller than the lower critical force $F_{c1} \approx 0.4$ all vortices are pinned to their pinning centers, i.e. vortices do not move at all, II) For forces between $F_{c1} \approx 0.4$ and $F_{c2} \approx 0.65$, only vortices in columns with half the number of pinning centers move (odd numbered columns, see Fig. 1). The Kagomé lattice is a hexagonal lattice with vacancies at the center of each the unit cell. Odd numbered columns have twice the number of pinning centers than even columns. III) For forces greater than the upper critical force $F_{c2}$ all vortices move. However, vortices in odd numbered channels move more slowly than vortices in channels with fewer pinning centers. This explains the existence of two peaks in the differential resistance. In all three regimes the diffusion is negligible.

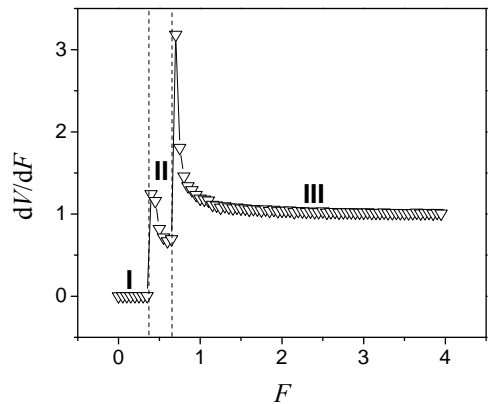

**Figure 2.** Differential resistance as a function of longitudinal transport force in a Kagomé pinning array.

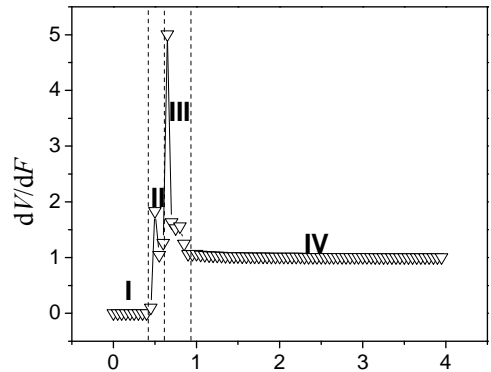

**Figure 3.** Differential resistance as a function of transversal transport force in Kagomé pinning array.

In Figure 3, we show the results for the Kagomé pinning array with transport forces along the transverse direction.

Different than in the longitudinal case, we observe four different dynamical regimes: I) The lower critical force is higher than in the longitudinal case, $F_{c1} \approx 0.45$, below which all vortices are pinned, II) For forces between $F_{c1}$ and $F_{c2} \approx 0.65$ vortices show a complex dynamical behavior as shown in Fig.4. Even rows have twice the number of pinning centers than odd rows, but different to the case with longitudinal force, alternate odd rows are unequal, i.e. the sequence of lattice sites with pinning centers and without pinning center is shifted. Vortices in rows with pinning vacancies (odd rows) move while vortices in rows with all vortices on pinning centers remain trapped. Vortices in odd rows do not move in a straight line but follow a sinusoidal trajectory from a given pinning site toward the nearest pinning site in an adjacent row and back toward the next pinning site of the same channel., III) For forces between $F_{c2}$ and $F_{c3} \approx 0,95$ vortices in rows with no vacancies start moving in addition to the movement of the vortices in rows with vacancies, as can be seen by the differential resistance in Fig. 3. However, vortex dynamics is a complicated channel network with interconnectivity, i.e., vortices can jump from one channel to another. This can be seen from the results for transversal diffusion. The analysis of the time average of structure factor shows that this dynamical regime resembles the smectic flow, observed in simulations with random pinning arrays [7]. This transition (which is denoted by the second peak on differential resistance) is not evident, because the analysis of structure factor shows that the flow changes from an ordered dynamical regime to a disordered one. Note the two peaks in the differential resistance are due to two different values for the driving force at which depinning of vortices occurs for incongruent channels. IV) For currents above $F_{c3}$, the movement of vortices is restricted to their corresponding channel preventing inter-channel vortex jumping. The differential resistance is constant and the diffusion coefficient in transverse direction (x) is very small.

## 4. CONCLUSION

In this work we report vortex dynamics simulations for a Kagomé pinning array for two directions of transport current. We find a remarkable smectic-like phase in a system with periodic pinning and an unusual multi-peak differential resistance, which reflects a complex dynamic behavior. This complex behavior is due to interstitial pinning i.e., rows with different pinning densities and topology have different values for the depinning force. Our results confirm an anisotropic behavior of the vortex system for forces in two mutually perpendicular directions in agreement with experiments [4].

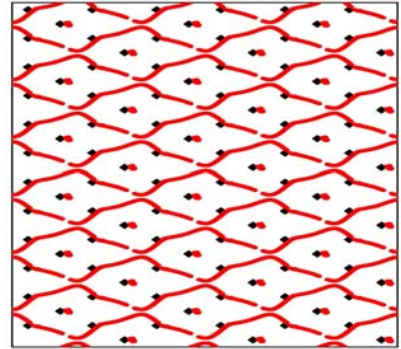

**Figure 4.** Vortex trajectory for Kagomé pinning array and transverse transport currents between $F_{c1}$ and $F_{c2}$.

## 5. AKNOWLEDGMENTS

This work was supported in part by FINEP and resources supplied by the Center for Scientific Computing (NCC/GridUNESP) of the São Paulo State University (UNESP) from Brazil.

## 6. REFERENCES

1. C. Reichhardt, C. J. Olson, and F. Nori, Phys. Rev. B **58**, 6534 (1998).
2. A. Castellanos, R. Wordenweber, G. Ockenfuss, A.V.D. Hart, K. Keck, Appl. Phys. Lett. **71**, 962 (1997).
3. M. F. Laguna, C. A. Balseiro, and D. Domingues, Phys. Rev. B **64**, 104505 (2001).
4. R. Cao, T.C. Wub, P.C. Kanga, J.C. Wua, T.J. Yangb, L. Hornga, Solid State Comm. **143**, 171 (2007)
5. J. D. Reis, P. A. Venegas, D. F. Mello, G. G. Cabrera, Physica C, **454,** 15 2007.
6. M. F. Laguna, C. A. Balseiro, D. Domínguez, and F. Nori, Phys. stat. sol. **230**, 499–503 (2002)
7. A. B. Kolton, D. Domínguez, N. Gronbech-Jensen, Phys. Rev. Lett. **83**, 3061 (1999).
8. G. Carneiro, Phys. Rev. B **57**, 6077 (1998).
9. C. Reichhardt, G. T. Zimányi, and N. Gronbech-Jensen, Phys. Rev. B **64**, 014501 (2001)
10. C. C. de Souza Silva, J. Van de Vondel, B. Y. Zhu, M. Morelle and V. V. Moshalkov, Phys. Rev. B **74**, 014507 (2006)
11. D. J. Morgan and J. B. Ketterson, Phys. Rev. Lett. **80**, 3614 (1998)
12. N. Gronbech-Jensen, Computer Phys. Communications **119**, 115 (1999).
13. C. Reichhardt and N. Gronbech-Jensen, Phys. Rev. B **63**, 54510 (2001)